\documentclass[pre,preprint,aps,amssymb,showpacs,superscriptaddress]{revtex4}

\begin{document}

\title[The Fermi-Pasta-Ulam paradox, Anderson Localization problem and the the generalized diffusion approach]
 {The Fermi-Pasta-Ulam paradox, Anderson Localization problem and the the generalized diffusion approach}

\author{V.N.~Kuzovkov}

\affiliation{Institute of Solid State Physics, University of Latvia,
8 Kengaraga Street, LV -- 1063 RIGA, Latvia}

\email{kuzovkov@latnet.lv}

\date{Received \today}

\begin{abstract}
The goal of this paper is two-fold. First, based on the
interpretation of a quantum tight-binding model in terms of a
classical Hamiltonian map, we consider the Anderson localization
(AL) problem as the Fermi-Pasta-Ulam (FPU) effect in a modified
dynamical system containing both stable and unstable (inverted)
modes. Delocalized states in the AL are analogous to the stable
quasi-periodic motion in FPU; whereas localized states are analogous
to thermalization, respectively. The second aim is to use the
classical Hamilton map for a simplified derivation of \textit{exact}
equations for the localization operator $H(z)$. The letter was
presented earlier [J.Phys.: Condens. Matter {\bf 14} (2002) 13777]
treating the AL as a generalized diffusion in a dynamical system. We
demonstrate that counter-intuitive results of our studies of the AL
are similar to the FPU counter-intuitivity.
\end{abstract}


\maketitle

\section{Introduction}

Half a century ago two celebrated papers were published temporally
close to each other which gave birth to two fundamental directions
of theoretical physics. In 1955 the \textit{Fermi-Pasta-Ulam}
(hereafter, FPU) \textit{paradox} was formulated
\cite{Fermi55,Focus05} which suggested the nonequipartition of
energy among normal modes of an anharmonic atomic chain. This
phenomenon is closely connected with the problems of ergocidity,
integrability, chaos and stability of motion
\cite{Focus05,Berman05}. A few years later, in 1958,
\textit{Anderson} \cite{Anderson58} suggested the possibility of
electron localization (AL) in a random system, provided that the
disorder is sufficiently large. This idea is one of the foundations
for the understanding the electronic properties of disordered
systems \cite{Abrahams01}. From a more general point of view, this
idea implies the absence of wave diffusion in a random medium as a
universal feature of stochastic processes. Unlike the AL problem,
where the stochasticity was explicitly introduced into consideration
through random potentials, the FPU considers the stochasticity as a
non-trivial effect in the dynamics of nonlinear systems, since a
strong chaos behaviour can be observed even in a system with several
degrees of freedom.

Despite the fact that the FPU paradox can be reformulated
analogously to the AL problem in terms of waves (ineraction of
normal modes which are characterized by their wave number $k$),
their conceptual similarity has not been noticed so far. In fact,
\textit{localization}, even when mentioned in the FPU literature, is
not associated with disorder; usually this means that the energy
initially placed in a low-frequency normal mode $k_0$ of the linear
problem stay almost completely locked within a few modes neighboring
the $k_0$ mode. The localization in $k$-space of normal modes
\cite{Flach05} or analogous energy localization in the FPU chain
\cite{Gershgorin05} are typically considered here. Sometimes the AL
and FPU are mentioned together, but as \textit{independent} and even
competing processes. In particular, energy transport in binary
isotropically disorded nonlinear FPU chains was considered
\cite{Snyder06} with the competition between localization (a
\textit{disorder} effect) and mode transitions (a
\textit{nonlinearity} effect). Notice also that the close connection
between the nonlinear dynamics and the AL was established for
nonlinear systems with a much smaller number of freedom degrees than
in the FPU problem. Thus, it was shown that the quantum kicked rotor
model \cite{Fishman82,Joos03} can be mapped onto the AL model. On
the other hand, the quantum kicked rotor serves as the starting
point of the systematic analysis of the quantum dynamics of
classically chaotic dynamical systems.

Our purpose is to establish a \textit{close connection} between the
two fundamental problems. That is, the AL is nothing else but the
FPU effect in a modified dynamical system with interacting normal
modes. Besides, the delocalized states in the AL problem are
analogous to the stable quasi-periodic motion (recurrence in the FPU
problem) and, respectively, the localized states are similar to the
\textit{thermalization} (motion instability) in the FPU. Our FPU
modification includes: (i) non-trivial change of the mode ensemble.
In the FPU without interactions all normal modes are stable. In
contrast, in the AL both stable and unstable (\textit{inverted})
modes also exist. (ii) a modified mode interaction: the
stochasticity in the AL is introduced directly, through random
forces \textit{linear} in coordinates and rather indirectly, through
nonlinear terms in the dynamical equations.

A critical comparison of the AL problem and the FPU paradox is the
more useful since the former turned out to be also a paradox.
Indeed, in recent years, the conflicting situation was established
here: (a) experimental results contradict the generally-accepted
theory \cite{Abrahams01,Kravchenko04}, whereas (b) the analytical
theory contradicts numerical simulations \cite{Markos06,Suslov06b}.
Despite such a clear conflict, the results of our \textit{exact}
analytical theory \cite{Kuzovkov02,Kuzovkov04,Kuzovkov06} are
considered as highly unexpected in the Anderson community
\cite{Comment,Reply,Suslov05,Suslov06c} since they contradict the
generally-accepted results of the scaling theory \cite{Abrahams79}
and numerical modeling \cite{Markos06}. In particular, doubts are
expressed about our conclusion on the existence of the
metal-insulator transition in the two dimensional (2D) disordered
system of noninteracting electrons (which does not contradict real
experimental data \cite{Abrahams01,Kravchenko04}) and that the
Anderson transition is of the first-order (localized and conducting
state co-exist). The latter statement is a clear example of a
counter-intuitive prediction. Appearance of highly unexpected
results is typical for such counter-intuitive (paradox) problems
such as the AL and FPU.

In this paper, we employ the main results of our analytical method
\cite{Kuzovkov02,Kuzovkov04,Kuzovkov06} and treat the AL as
generalized diffusion in a dynamical system. The random forces
impose random walk amplitudes (the dynamics is bounded in the phase
space for unperturbed system) which can lead to the diffusion
divergence. The diffusion problem with random forces linear in the
coordinates can be exactly solved. Moreover, the diffusion concept
permits to connect the AL and FPU problems. As is well-known
\cite{Joos03}, the equations for nonlinear dynamical systems can
under certain conditions describe pseudo-random walks which leads to
the diffusion behaviour and diffusion increase in mean energy (i.e.
divergence).

We shall employ also the interpretation of the Schr\"odinger
equation (quantum tight-binding model) in terms of the classical
Hamiltonian map \cite{Izrailev98,Izrailev99,Tessieri00}. As a
result, the problem can be reformulated in terms of interacting mode
dynamics which opens the opportunity for the detailed comparison of
AL with FPU. It should be stressed that ideas
\cite{Izrailev98,Izrailev99,Tessieri00} allow to simplify
considerably the mathematical formalism used earlier
\cite{Kuzovkov02,Kuzovkov04,Kuzovkov06}. This is important in the
light of the recent criticism \cite{Comment,Suslov05} that the
\textit{engineering language} (signal theory) of the mentioned
formalism (input and output signals, filter function, etc.) is new
for the AL community. In this paper we suggest a compact derivation
of the main feature of the disordered system - the localization
operator $H(z)$ \cite{Kuzovkov06}.

The structure of the paper is as follows. In Section \ref{Cauchy} we
explain how the one-dimensional tight-binding model with diagonal
disorder can be presented in terms of the classical two-dimensional
Hamiltonian map for normal or inverted oscillators. In Section
\ref{Classic} we present the equations for arbitrary dimensions. We
show that delocalized states, in general, correspond to the
statistically bound trajectories, whereas localized states to
unbound trajectories, respectively. The trajectory type depends on
the excited mode (normal or inverted). We show in Section
\ref{Equations} that statistically unbound trajectories can be
treated in terms of a \textit{generalized diffusion}, with exactly
predicable properties. As a result, we arrive in Section
\ref{Operator} at the definition of the localization operator
$H(z)$. In Section \ref{FPU} a possible comparison of the AL and FPU
problems is discussed. It is shown that the AL treatment in terms of
the classical Hamiltonian map results in a paradox demonstrating the
counter-intuitive nature of this problem. The results
\cite{Kuzovkov02,Kuzovkov04,Kuzovkov06} are compared with those for
the FPU and their detailed similarities are analyzed.

\section{Anderson localization and classical Hamiltonian
map }
\subsection{Cauchy problem and  classical Hamiltonian
map for the one-dimensional case}\label{Cauchy}

As is well known, in order to determine the Lyapunov exponent
$\gamma$ (which is the inverse of the localization length,
$\xi=1/\gamma$) and the phase diagram (the areas of the localized
and delocalized states), the Cauchy problem with fixed initial
conditions has to be solved
\cite{Markos06,Kuzovkov02,Suslov05,Molinari92}. For illustration,
the 1D Schr\"odinger equation
\begin{equation}
\psi_{n+1}+\psi_{n-1}=E\psi_n -\varepsilon_n \psi_n
\end{equation}
with random potentials $\varepsilon_n$ can be presented as a
recursive relation
\begin{equation}\label{eq02}
\psi_{n+1}=E\psi_n-\psi_{n-1} -\varepsilon_n \psi_n .
\end{equation}
The treatment of one of the spatial coordinates as a temporal
variable (discrete time $n$) is a standard approach in chaos theory
\cite{Zaslavsky} which opens the way to the dynamical
interpretation. Taking into account the so-called \textit{causality
principle} \cite{Kuzovkov02,Kuzovkov06}, the latter equation permits
an exact stochastic analysis. Indeed, it is easy to see that in the
Cauchy problem (with fixed initial conditions for $\psi_1$ and
$\psi_0$), eq.(\ref{eq02}),  $\psi_2$ is a function of
$\varepsilon_1$, $\psi_3$ is a function of
$\varepsilon_2$,$\varepsilon_1$,  etc (a \textit{causality}). That
is both amplitudes $\psi_n$ and $\psi_{n-1}$ on the rhs of
eq.(\ref{eq02}) are statistically independent of $\varepsilon_n$ and
can be averaged separately (causality principle):
\begin{eqnarray} \label{eq222}
\langle\psi_{n+1}\rangle=(E-\langle\varepsilon_n\rangle)\langle\psi_{n}\rangle-\langle\psi_{n-1}\rangle
,
\end{eqnarray}
\begin{eqnarray} \label{eq333}
\langle\psi^2_{n+1}\rangle=(E^2-2E\langle\varepsilon_n\rangle+
\langle\varepsilon^2_n\rangle)\langle\psi^2_{n}\rangle
- \\
\nonumber -2(E-\langle\varepsilon_n\rangle)
\langle\psi_n\psi_{n-1}\rangle+\langle\psi^2_{n-1}\rangle.
\end{eqnarray}
The causality principle can be used only (a) for the recursive
relation for the Cauchy problem and (b) when on-site potentials are
independently distributed ($\langle \varepsilon_n
\varepsilon_{n^{\prime}} \rangle=\sigma^2 \delta_{n,n^{\prime}}$)
but not for the Dirichlet problem, correlated potentials, etc.

To study the origin of localized/delocalized states, we use a simple
approach \cite{Izrailev98,Izrailev99,Tessieri00} based on the
interpretation of a quantum tight-binding model, eq.(\ref{eq02}),
with diagonal disorder in terms of the classical two-dimensional
Hamiltonian map. The difference equation (\ref{eq02}) is reduced to
a discrete transform with a simple physical interpretation. Let us
write the second-order equation as a set of two first-order
equations. Assume $q_n=\psi_n$ and $p_n=\psi_{n+1}-\psi_n$ for $E
\geq 0$; whereas $q_n=-\psi_n$ and $p_n=-(\psi_{n+1}-\psi_n)$ as  $E
< 0$. The obtained equation set reads
\begin{eqnarray}\label{p}
p_{n+1}=p_n-\omega^2 q_n-\varepsilon_n q_n ,\\
q_{n+1}=q_n+p_{n+1} ,\label{q}
\end{eqnarray}
where $\omega^2=2-|E|$.

As is shown \cite{Zaslavsky}, the discrete transform (\ref{p}),
(\ref{q}) can be connected with the equivalent differential equation
of the Hamilton dynamics with the Hamiltonian (kicked oscillator)
\begin{eqnarray} \label{osc}
\mathcal{H}=\frac{p^2}{2}+\frac{\omega^2q^2}{2}+\frac{\varepsilon(t)q^2}{2}\sum_n
\delta(t-n\Delta t) .
\end{eqnarray}
It defines the system with the unperturbed Hamiltonian for an
oscillator affected by a periodic sequence of a kicks
($\delta$-pulses) with the period $\Delta t$. Discrete transform
arises when magnitudes of coordinate and impulse are considered with
a discrete time increment $\Delta t$ and correspondingly with a
discrete time index $n$ (discrete time).

Dependent on the $\omega^2$ sign, there exist two cases. For
$|E|<2$, corresponding to the solutions inside the band in the
unperturbed system, we obtain a normal (stable) oscillator with
$\omega^2>0$. On the contrary, the energies outside the band,
$|E|\geq 2$, are associated with inverted (unstable) oscillators
with $\omega^2 \leq 0$. Without disorder $\varepsilon_n \equiv 0$
and delocalized quantum states correspond to a normal oscillator and
trajectories bound in the classical $(p,q)$ space as $n\rightarrow
\infty$ \cite{Izrailev99}. Simple characteristics of the dynamical
system $q_n^2$ or $p_n^2$ are bound, respectively. Contrary, an
inverted oscillator describes non-physical solutions outside the
band which are now unbound in the classical phase space; the $q_n^2$
magnitude is divergent as $n\rightarrow \infty$.

When disorder is introduced, the situation changes qualitatively.
Random kicks for a normal oscillator lead to the random amplitude
walks. The oscillatory motion remains since the average $\langle q_n
\rangle$ remains bound as $n\rightarrow \infty$. For a random
amplitude walk long trajectories in the classical phase space
$(p,q)$  are possible which can be treated as diffusion
\cite{Kuzovkov06}. This is characterized by a typical parameter
divergence: $\langle q_n^2 \rangle \rightarrow \infty$ as
$n\rightarrow \infty$. Since these trajectories correspond in 1D to
the localized states, these can be considered as statistically
unbound. In the case of the inverted oscillator the amplitude
increases exponentially (in the model with Hamiltonian
eq.(\ref{osc})) between successive kicks, but the force linear in
coordinates is able to change a coordinate sign.

Use of the diffusion terminology is quite justified here. Indeed, to
detect the diffusion, it is sufficient to demonstrate the divergence
of the second moment of the amplitude $q_n$ and to establish its
time-dependence, the function $f(n)$ in $\left \langle q^2_n
\right\rangle =f(n)$. The divergence of the second moment defines
the conditions of the diffusion appearance. For a \textit{normal
diffusion} the mean square displacement is linear in time, $\left
\langle q^2_n \right\rangle \propto n$. The notion of an
\textit{anomalous diffusion} \cite{Sokolov05,Klages08} derives from
the fact that the mean square displacement may be anomalously
diffusive, $\left \langle q^2_n \right\rangle \propto n^{\alpha}$
($\alpha\neq 1$), i.e. nonlinear in time (power-law divergence). The
quantity $\left \langle q^2_n \right\rangle = \left \langle \psi^2_n
\right\rangle$ in eq.(\ref{eq02}) was calculated analytically
\cite{Kuzovkov02,Molinari92}: $\left \langle q^2_n \right\rangle
\propto \exp(2\gamma n)$ with $\gamma \geq 0$ for an arbitrary $E$
value, provided $\sigma>0$. In this case the  \textit{generalized
diffusion} (exponential divergence) takes place. The appearance of
the localization in the approach based on eq. (\ref{eq02}) is
equivalent to the appearance of diffusion. Respectively, the
well-known statement that in one dimension all states are localized
at any level of disorder, is equivalent to the statement on the
diffusion character of all solutions of  eq.(\ref{eq02}) in 1D for
$\sigma
>0$.

\subsection{Classical Hamiltonian
map for D-dimensional case} \label{Classic}

The above-discussed statement of the problem can be naturally
generalized for an arbitrary dimension D; this idea was mentioned
but not realized in Ref. \cite{Tessieri00}. The phase diagram of the
system with metal-insulator transition should be obtained in the
thermodynamical limit (the infinite system). Let us consider the
semi-infinite system, or an infinite system with a boundary, where
the index $n \equiv m_D\geq 0$, but all $m_j\in(-\infty,\infty)$,
$j=1,2,\dots,p$, with $p=D-1$. We combine indices  in the form of a
vector $\mathbf{m}=\{m_1,m_2,\dots,m_p\}$. The boundary which is the
layer $n=0$ defines the preferential direction (the axis $n$).

The Schr\"odinger equation can be rewritten as a recursion equation
(in terms of the discrete-time $n$)
\begin{equation} \label{recursion DN}
\psi _{n+1,\mathbf{m}}=(E-\varepsilon _{n,\mathbf{m}}) \psi
_{n,\mathbf{m}}-\psi_{n-1,\mathbf{m}} - \sum_{\mathbf{m^{\prime}}}
\psi_{n,\mathbf{m^{\prime}}} .
\end{equation}
Summation over $\mathbf{m^{\prime}}$ runs over the nearest
neighbours of the site $\mathbf{m}$. The on-site potentials $
\varepsilon_{n,m} $ are independently and identically distributed.
We assume hereafter existence of the two first moments,
$\left\langle \varepsilon _{n,\mathbf{m}} \right\rangle =0$ and
$\left\langle \varepsilon _{n,\mathbf{m}}^2\right\rangle =\sigma
^2$, where the parameter $\sigma$ characterizes the disorder level.

Let us perform the Fourier transform:
\begin{eqnarray}
\varepsilon_n(\mathbf{k})=\sum_{\mathbf{m}} \varepsilon_{n,\mathbf{m}}e^{i\mathbf{k}\mathbf{m}} ,\\
\psi_n(\mathbf{k})=\sum_{\mathbf{m}}
\psi_{n,\mathbf{m}}e^{i\mathbf{k}\mathbf{m}} .
\end{eqnarray}
The relation for similar random quantities, , $\langle
\varepsilon_{n,\mathbf{m}} \rangle =0$, $\langle
\varepsilon_{n,\mathbf{m}}\varepsilon_{n^{\prime},\mathbf{m}^{\prime}}
\rangle =\sigma^2
\delta_{n,n^{\prime}}\delta_{\mathbf{m},\mathbf{m}^{\prime}}$, leads
to
\begin{eqnarray}
\langle \varepsilon_n(\mathbf{k}) \rangle =0 ,\\\label{sig} \langle
\varepsilon_n(\mathbf{k})\varepsilon^{*}_{n^{\prime}}(\mathbf{k}^{\prime})
\rangle =(2\pi)^p \sigma^2 \delta_{n,n^{\prime}}\delta
(\mathbf{k}-\mathbf{k}^{\prime}) .
\end{eqnarray}
As a result of the  Fourier transform, the Schr\"odinger equation
(\ref{recursion DN}) transforms into the equation set for the mode
dynamics, enumerated by the index $\mathbf{k}$,
\begin{equation}\label{Sch}
\psi_{n+1}(\mathbf{k})=\mathcal{L}(\mathbf{k})\psi_n(\mathbf{k})-\psi_{n-1}(\mathbf{k})-\int
\frac{d^p\mathbf{k}_1}{(2\pi)^p}
\varepsilon_n(\mathbf{k}-\mathbf{k}_1)\psi_n(\mathbf{k}_1) .
\end{equation}
Here
\begin{equation}
\mathcal{L}(\mathbf{k})=E-2\sum_{j=1}^{p=D-1}\cos(k_j) ,
\end{equation}
with fixed initial conditions $\psi_0(\mathbf{k})$ and
$\psi_1(\mathbf{k})$.

A comparison of eq.(\ref{Sch}) with its 1D analog, eq.(\ref{eq02}),
demonstrates that the dynamics of the multi-dimensional system can
be reduced to the dynamics of the multi-oscillatory system. The
frequencies of these oscillators are defined by the relation
$\omega(\mathbf{k})^2=2-|\mathcal{L}(\mathbf{k})|$. That is, one can
distinguish, as before, normal oscillators with
$|\mathcal{L}(\mathbf{k})| <2 $ and inverted oscillators with
$|\mathcal{L}(\mathbf{k})| \geq 2$. As we have shown
\cite{Kuzovkov06}, the condition $|\mathcal{L}(\mathbf{k})| <2 $
without perturbation corresponds  to delocalized states inside the
band; and contrary, $|\mathcal{L}(\mathbf{k})| \geq 2$ corresponds
to the solution outside the band. In other words, the connection
between the classical oscillator type (normal or inverted) and the
quantum-mechanical solution remains also for arbitrary dimensions.
However, there is also an important difference between the 1D and ND
systems: in the former case the energy magnitude determines uniquely
the oscillator type. The terms with a random force describe only the
oscillator's stochastic self-interaction, due to its linear
dependence, $-\varepsilon_nq_n$ in eq.(\ref{p}), the oscillator
cannot be stopped, $q=p=0$.

In contrast, in the multi-dimensional case the energy $E$ no longer
determines uniquely the system's state, the fixed initial conditions
$\psi_0(\mathbf{k})$ and $\psi_1(\mathbf{k})$ define simultaneously
the type and number of initially excited oscillators, among which
can be found both normal and inverted oscillators. The integral in
eq.(\ref{Sch}) corresponding to a random force, describes now the
stochastic interaction between oscillators. That is, knowledge of
the 1D system is \textit{not} sufficient for the description of
multi-dimensional systems; as we demonstrate below, fundamentally
new effects arise here.

Using the \textit{causality principle} \cite{Kuzovkov02,Kuzovkov06},
the equation for the first momentum of a random amplitude $\langle
\psi_n(\mathbf{k}) \rangle$ is quite trivial:
\begin{equation}\label{Sch1}
\langle \psi_{n+1}(\mathbf{k}) \rangle
=\mathcal{L}(\mathbf{k})\langle
\psi_n(\mathbf{k})\rangle-\langle\psi_{n-1}(\mathbf{k})\rangle .
\end{equation}
It is easy to see that for unstable modes,
$|\mathcal{L}(\mathbf{k})| \geq 2$, even the first moments are
divergent, $|\langle \psi_n(\mathbf{k})\rangle| \rightarrow \infty$,
as $n \rightarrow \infty$. Its analog in the classical phase space
$(p,q)$ corresponds to unbound trajectories.

Since we associate the appearance of localized states with unbound
trajectories for classical oscillators, it is easy to formulate the
necessary (but not sufficient) condition for the appearance with
disorder of the delocalized states: the initial conditions should
correspond to excitation of normal oscillators only, i.e. the
amplitudes $\psi_0(\mathbf{k})$ and $\psi_1(\mathbf{k})$ are nonzero
only for modes with $|\mathcal{L}(\mathbf{k})| < 2$. This coincides
with the statement \cite{Kuzovkov06} based on different ideas. Under
this condition the dynamics of the first moments is bound for all
modes, $|\langle \psi_n(\mathbf{k})\rangle| <\infty$  as $n
\rightarrow \infty$. However, this condition is not sufficient,
since the localized states as it was illustrated for the 1D case,
correspond in general to \textit{statistically unbound
trajectories}, $\langle |\psi_n(\mathbf{k})|^2 \rangle \rightarrow
\infty$ as $n \rightarrow \infty$. That is, the search for the
sufficient condition for the excistence of the delocalized states is
reduced to the solution of equations for the \textit{second moments}
of the random amplitudes \cite{Kuzovkov02,Kuzovkov06}.

\subsection{Equations for second moments}\label{Equations}

Divergence of the second moments is a typical diffusion behaviour;
an observation of such a diffusion dynamics indicates directly the
presence of the localized states, and vice versa. An easy criterium
of a diffusion is the behaviour of the squared coordinate for all
oscillators
\begin{equation}\label{Un}
U_n=\int \frac{d^p\mathbf{k}}{(2\pi)^p}
\langle|\psi_n(\mathbf{k})|^2\rangle .
\end{equation}
To detect the diffusion, it is sufficient to demonstrate divergence
of the function $U_n$ as $n \rightarrow \infty$.

Let us define the second moments by the relations:
\begin{eqnarray}
x_n(\mathbf{k}) =\langle|\psi_n(\mathbf{k})|^2\rangle
=\langle\psi_n(\mathbf{k}) \psi^{*}_n(\mathbf{k})\rangle ,
\\
y_n(\mathbf{k}) =\frac{1}{2}[\langle\psi_n(\mathbf{k})
\psi^{*}_{n-1}(\mathbf{k})+\psi^{*}_n(\mathbf{k})
\psi_{n-1}(\mathbf{k})\rangle ] .
\end{eqnarray}
One gets
\begin{equation}\label{U_n}
U_n=\int \frac{d^p\mathbf{k}}{(2\pi)^p} x_n(\mathbf{k}) .
\end{equation}
Using the causality principle (see details in
Ref.\cite{Kuzovkov02,Kuzovkov06})) for eq.(\ref{Sch}), one gets
immediately for the non-zero average quantities:
\begin{eqnarray}
\langle|\psi_{n+1}(\mathbf{k})|^2
\rangle=\mathcal{L}^2(\mathbf{k})\langle|\psi_n(\mathbf{k})|^2
\rangle+\langle|\psi_{n-1}(\mathbf{k})|^2 \rangle\\
- \nonumber \mathcal{L}(\mathbf{k})\langle
[\psi_n(\mathbf{k})\psi^{*}_{n-1}(\mathbf{k})+ \psi^{*}_n(\mathbf{k})\psi_{n-1}(\mathbf{k})\rangle \\
\nonumber + \int \int \frac{d^p\mathbf{k}_1}{(2\pi)^p}
\frac{d^p\mathbf{k}_2}{(2\pi)^p}
\langle\varepsilon_n(\mathbf{k}-\mathbf{k}_1)\varepsilon^{*}_n(\mathbf{k}-\mathbf{k}_2)\rangle
\langle\psi_n(\mathbf{k}_1)\psi^{*}_n(\mathbf{k}_2)\rangle .
\end{eqnarray}
Taking into account properties of random potentials, eq.(\ref{sig}),
one gets for для $n\geq 1$:
\begin{eqnarray}\label{x}
x_{n+1}(\mathbf{k})=\mathcal{L}^2(\mathbf{k})x_n(\mathbf{k})+x_{n-1}(\mathbf{k})-2\mathcal{L}(\mathbf{k})y_n(\mathbf{k})
+\sigma^2 U_n .
\end{eqnarray}
Notice that the last term in eq.(\ref{x}) does not depend on
$\mathbf{k}$, i.e. the noise equally affects \textit{all} modes. The
noise intensity is described by $\sigma^2 U_n$, where $U_n$ was
defined in eq.(\ref{U_n}). Analogously, the complementary equation
is derived
\begin{eqnarray}\label{y}
y_{n+1}(\mathbf{k})+y_{n}(\mathbf{k})=\mathcal{L}(\mathbf{k})x_n(\mathbf{k})
.
\end{eqnarray}
Let us perform now the Z-transform:
\begin{eqnarray}
X(z,\mathbf{k})=\sum_{n=1}^{\infty} \frac{x_n(\mathbf{k})}{z^n} ,\\
Y(z,\mathbf{k})=\sum_{n=1}^{\infty} \frac{y_n(\mathbf{k})}{z^n} ,\\
U(z)=\sum_{n=1}^{\infty} \frac{U_n}{z^n} .
\end{eqnarray}
Takin into account eq.(\ref{U_n}), one gets easily
\begin{equation}\label{U(z)}
U(z)=\int \frac{d^p\mathbf{k}}{(2\pi)^p} X(z,\mathbf{k}) .
\end{equation}
The Z-transform of eqs.(\ref{x}), (\ref{y}) leads to the relations
containing the initial conditions
\begin{eqnarray}
[z-z^{-1}-\mathcal{L}^2(\mathbf{k})]X(z,\mathbf{k})+2\mathcal{L}(\mathbf{k})Y(z,\mathbf{k})=\\
\nonumber x_1(\mathbf{k})+z^{-1}x_0(\mathbf{k})+\sigma^2U(z)
,\\
(z+1)Y(z,\mathbf{k})-\mathcal{L}(\mathbf{k})X(z,\mathbf{k})=y_1(\mathbf{k})
.
\end{eqnarray}
It is easy to find that
\begin{eqnarray}\label{X(z)}
\frac{(z-1)}{(z+1)}[(z+1)^2/z-\mathcal{L}^2(\mathbf{k})]X(z,\mathbf{k})=\lambda(z,\mathbf{k})+\sigma^2U(z)
,
\end{eqnarray}
where
$\lambda(z,\mathbf{k})=2\mathcal{L}(\mathbf{k})y_1(\mathbf{k})/(z+1)+x_0(\mathbf{k})+x_1(\mathbf{k})/z$.
Taking into account the definition (\ref{U(z)}), eq.(\ref{X(z)})  is
the integral equation, but with a simple structure.

The initial conditions for moments are easily expressed through
$\psi_0(\mathbf{k})$, $\psi_1(\mathbf{k})$.  The trivial result is
that if any mode with the wave vector $\mathbf{k}_0$ enters the
initial conditions (i.e. at least one of the two quantities
$\psi_0(\mathbf{k})$, $\psi_1(\mathbf{k})$ is non-zero), the
quantity $\lambda(z,\mathbf{k}_0)$ is also non-zero for this mode.
This will be used in the further analysis.

\subsection{Localization operator}\label{Operator}

Let us assume for the beginning that $\sigma=0$, i.e. there is no
disorder in eq.(\ref{X(z)}), and use the relevant solution
$X^{(0)}(z,\mathbf{k})$  for the calculation of the squared
coordinate, eq.(\ref{U(z)}). Thus, one gets
\begin{eqnarray}\label{Uo}
U^{(0)}(z)=\frac{(z+1)}{(z-1)}\int
\frac{d^p\mathbf{k}}{(2\pi)^p}\frac{\lambda(z,\mathbf{k})}{[(z+1)^2/z-\mathcal{L}^2(\mathbf{k})]}
.
\end{eqnarray}
For $\sigma \neq 0$  the solution reads
\begin{eqnarray}\label{HU}
U(z)=H(z)U^{(0)}(z) .
\end{eqnarray}
Here $H(z)$ is the localization operator
\begin{eqnarray} \label{H}
\frac{1}{H(z)}=1- \sigma^2\frac{(z+1)}{(z-1)}\int
\frac{d^p\mathbf{k}}{(2\pi)^p}\frac{1}{[(z+1)^2/z-\mathcal{L}^2(\mathbf{k})]}
.
\end{eqnarray}
Using the convolution property for the Z-transform
\cite{Kuzovkov02,Kuzovkov06}, one gets
\begin{eqnarray}\label{conv}
U_n=\sum_{l=1}^{n}U^{(0)}_{l} h_{n-l} ,
\end{eqnarray}
where $h_n$ is the result of the inverse Z-transform of the
localization operator $H(z)$.

As was mentioned, the necessary condition for the existence in the
presence of disorder of the delocalized states is the presence at
the initial time of normal (stable) modes only. In this case
$U^{(0)}_n$ corresponds to the stable dynamics of the unperturbed
problem and thus is bound in the index $n$: $U^{(0)}_n < \infty$ as
$n \rightarrow \infty$. In its turn, the sufficient condition for
the delocalized states is the \textit{absence} of the (diffusion)
$U_n$ divergence: $U_n < \infty$ as $n \rightarrow \infty$. As soon
as $U_n \rightarrow \infty$, as $n \rightarrow \infty$, this
indicates the localized states.

It is easy to notice that the convergence or divergence of $U_n$ is
not dependent on the properties of the unperturbed solutions. The
problem is reduced to a study of the asymptotic behaviour ($n
\rightarrow \infty$) of the $h_n$ coefficients in the linear
transformation, eq.(\ref{conv}) \cite{Kuzovkov02}. This does not
need calculation of the coefficients $h_n$ but, by the means of
analytical methods, analysis of the localization operator $H(z)$  as
a function of the complex variable $z$. Therefore, the physical
problem of the localized/delocalized states is reduced to the
mathematical search for the poles of the function $H(z)$ of the
complex variable \cite{Kuzovkov02,Kuzovkov04}. We do not go into
details here; in particular, the diagrammic technique of the search
for the poles was discussed in Appendix of Ref.\cite{Kuzovkov04}.

In particular, it was shown \cite{Kuzovkov02,Kuzovkov04}  that the
localization operator $H(z)$ is a non-analytic function of the
complex variable $z$. The unit circle $|z|=1$ divides the complex
plane into two analytic domains: the interior and exterior of the
unit circle. The inverse Z-transform is quite generally defined via
countour integrals in the complex plane
\begin{equation}\label{inverse}
h_n=\frac{1}{2\pi i}\oint H(z)z^n \frac{dz}{z} .
\end{equation}
This definition is only possible in an analytic domain and does not
always represent a solution which can be physically interpreted
\cite{Kuzovkov02,Kuzovkov04}. In this way, multiple solutions can
result in the formal analysis of the problem. The first solution
$H_{+}(z)$ describes the localized states. It is defined outside the
unit circle and always exists. The second solution $H_{-}(z)$
describes delocalized states. It is defined inside the unit circle.
The coexistence of the two solutions (if any) was physically
interpreted  \cite{Kuzovkov02,Kuzovkov04,Kuzovkov06} as the
coexistence of two \textit{phases} -- an insulating and a metallic
one.

Notice that the same result can be obtained using, instead of the
total squared coordinate, eq.(\ref{Un}), the total squared momentum
as a criterium of the diffusion dynamics
\begin{equation}\label{Vn}
V_n=\int \frac{d^p\mathbf{k}}{(2\pi)^p}
\langle|\psi_{n+1}(\mathbf{k})-\psi_n(\mathbf{k})|^2\rangle .
\end{equation}
After simple transformations one gets
\begin{eqnarray}
V(z)=H(z)V^{(0)}(z) ,
\end{eqnarray}
whose structure is similar to that of eq.(\ref{HU}), the definition
of the localization operator $H(z)$ also retains, eq.(\ref{H}),
whereas $V^{(0)}(z)$ is the squared momentum of the unperturbed
system.

The expression for the localization operator $H(z)$ has been derived
by us earlier \cite{Kuzovkov02,Kuzovkov04}. However, the
nontraditional \textit{engineering language} was used (such as
input/output signals, filter function, etc). As we demonstrated
above, the derivation can be considerably simplified, since in the
system of interacting modes the asymptotic behaviour of the total
squared coordinate $U_n$ serves as a natural indicator of the
presence/absence of diffusion in the system's dynamics. Therefore,
reformulation of the Schr\"odinger equation (quantum tight-binding
model) in terms of the classical Hamiltonian map permits to retain
the basic definitions of the alternative approach
\cite{Kuzovkov02,Kuzovkov04,Kuzovkov06}, but opens the additional
possibility of a new interpretation of the results obtained earlier,
which we discuss in the next Section.

\section{The FPU problem  vs the AL problem}\label{FPU}

\subsection{Stability and thermalization}

It is generally believed that an increase in space dimension greatly
increases the system's stability with respect to disorder. Thus,
there is without doubt the presence of a metal-insulator transition
in the 3D case. It is believed that the effect of statistical
fluctuations changes the regime at $D=4$ \cite{Lee85,Kunz83}; no
phase transitions are expected for $D>4$. The 2D system marks the
\textit{borderline} between high and low dimensions \cite{Rice97}.
However, all these conclusions derive from the phenomenological
scaling theory of localization \cite{Abrahams79}. The alternative
point of view with different classification of high and low
dimensions has been presented by us in Ref.\cite{Kuzovkov04}. In
general, this confirms the system stability in higher dimensions.

An advantage of our interpretation of the AL in terms of classical
dynamics of stochastically interacting oscillators is that it makes
the statement on the relation between system stability and space
dimension to be not so obvious. Indeed, for the energy range inside
the old band $|E|\leq 2D$, where normal modes certainly exist,
simultaneously the inverted modes with $|\mathcal{L}(\mathbf{k})|
\geq 2$ are also always possible. Change of the space dimensions D
affects the weights of these states, but their \textit{complete}
disappearance is impossible. That is, this problem can be resolved
only by means of the exact analytical solution.

On the other hand, the \textit{instability} mechanism for
multi-dimensional systems ($D>1$) in terms of oscillators is quite
obvious. Even assuming that at the beginning only normal modes are
excited, stochastic interaction inavoidably excites also
neighboring, in particular, inverted modes. In other words,
\textit{thermalization} of all modes takes place. It is important to
stress that \textit{all} modes -- normal and inverted ones --
contribute to the localization operator $H(z)$, irrespective of
which modes were excited in the beginning. The inverted mode
dynamics in the classical phase space $(p,q)$ corresponds to unbound
trajectories which do not correspond to the delocalized states.
However, an immediate conclusion suggests itself that all solutions
of the dynamical problem -- independent of the space dimension and
disorder level -- correspond only to the localized states. Such a
paradox conclusion demonstrates clearly that the problem under
consideration is counter-intuitive.

In our opinion, the \textit{AL paradox} has much in common with the
FPU problem \cite{Fermi55,Berman05}. To show their close similarity,
let us summarize here the main results.

Numerical simulations of a chain of harmonic oscillators coupled
with a quadratic or cubic nonlinearity show that energy, initially
placed in a low-frequency normal mode of the linear problem stay
almost completely locked \textit{within a few neighbor modes} (or
\textit{quasi-modes} \cite{Berman05}), instead of being distributed
among \textit{all modes of the system}. Recurrence of energy to the
originally excited mode is also observed.  The nonlinear effects are
significant and cannot be neglected.

Two alternative explanations of the FPU paradox were suggested
\cite{Berman05}):  the integrability of nonlinear equations and
dynamical (deterministic) chaos. The second approach points to the
existence of a \textit{stochasticity threshold} in the FPU problem.
If the nonlinearity is below a stochasticity threshold, the dynamics
of the system remains similar to the one of the unperturbed  system
for large time scales. For a strong nonlinearity the overlap of
nonlinear resonances leads to a strong dynamical chaos, destroying
the FPU effect. Namely in this case the intuitive thermalization
occurs.

Therefore, in the standard FPU statement nonlinear effects play a
\textit{key role} being responsible for stochasticity: the motion of
a nonlinear dynamical system even with few degrees of freedom can
exhibit chaotic behavior. On the other hand, nonlinearity of the
dynamics equations prevents their analytical analysis. A number of
questions still remain open. In particular, the main results are
obtained for systems with few degrees of freedom, the behavior of
the system is not known in the thermodynamic limit, when the number
of degrees of freedom goes to infinity.

In our interpretation of the Schr\"odinger equation in terms of the
classical Hamiltonian map, stochasticity is introduced through
on-site potentials, which are random variables. Therefore, there is
no longer the need to solve nonlinear equations. For the Cauchy
problem with fixed initial conditions, where the \textit{causality
principle} can be applied  \cite{Kuzovkov02,Kuzovkov06}, this
problem can be solved \textit{exactly analytically} (see Section
\ref{Operator}), both for finite-size systems and its thermodynamic
limit. Unlike the FPU problem with counter-intuitive solution, the
classical interpretation of the AL problem is more complicated due
to the unobvious dynamics of the inverted modes.

We will consider below the delocalized states in the quantum
mechanical problem as statistically stable quasi-periodic motion, in
terms of the classical Hamiltonian map, with excitations spreading
only over a few neighbor modes (of the initially excited normal
modes). Such a dynamics is bound. The localized states are
interpreted respectively as diffusion dynamics with statistically
unbound trajectories. In the FPU problem this means thermalization
with an excitation spreading over \textit{all} modes.

\subsection{Stochasticity threshold vs disorder threshold}

As was noticed in Ref.\cite{Berman05},  the existence of a
\textit{stochasticity threshold} in the FPU is the nonlinear effect.
For a strong nonlinearity the overlap of nonlinear resonances leads
to a dynamical chaos, destroying the FPU recurrence.  As a result,
fast convergence to thermal equilibrium arises. The problem of the
nonlinearity prevents its detailed analytical study. The
stochasticity threshold depends on the type of nonlinearity, space
dimension; the thermodynamic limit is also unclear. Compared to this
situation, the dynamical interpretation of the Schr\"odinger
equation has obvious advantages permitting an exact analytical
solution. It is expected that in the stochastic AL problem the
relevant stochasticity threshold transforms into the
\textit{disorder threshold}.

When comparing the two problems, one has to keep in mind that the AL
was formulated as a statistical problem with an ensemble of random
potential realizations. Thus,  it is convenient to reformulate the
FPU as a statistical problem. Introducing the stochasticity
parameter $K$ (nonlinearity coefficient \cite{Zaslavsky}), the type
of dynamical trajectories is defined by the initial conditions. In
other words, in the nonlinear dynamic problem the stochasticity
parameter is not a unique factor determining the dynamics. For
example, the behavior of the FPU system depends strongly on whether
low- or high-frequency modes are initially excited \cite{Berman05}.
The number of excited modes seems also to be important for the
dynamics.

As is well known, in such problems the phase space is divided into
regions with qualitatively different types of motion, and these
regions are separated by barriers. If initial conditions were chosen
in the region corresponding to a stable quasi-periodic motion, this
dynamics corresponds to the recurrent behavior as in the FPU
experiment and is classified as the dynamics below the stochasticity
threshold. In another region an instability of motion exists for a
wide range of the initial conditions (the dynamics above the
stochasticity threshold). The value of the stochasticity parameter
$K$ determines the borders of these regions.

Let us now define some domain in phase space and an ensemble of the
initial conditions therein. If for a given $K$ a whole domain chosen
falls into some region, all trajectories in this ensemble reveal the
same dynamics -- stable quasi-periodic motion or dynamical chaos
(single phase domain) -- otherwise an ensemble reveals trajectories
of different kinds (two-phase domain). In the latter case trivial
\textit{co-existence} of the two phases (or two \textit{dynamics},
in a statistical sense) takes place. Change of the stochasticity
parameter $K$ in a given ensemble leads to the phase transitions. It
is easy to show that in our statistical problem phase transitions
from single-phase to hetero-phase (phase co-existence) dominate, and
vice versa, since these transitions correspond to the boundary
motion of the regions. Transitions from one single-phase system to
another single-phase are also possible since variation in the $K$
parameter in the phase space can induce spontaneous creation of new
regions with a different dynamics.

The Schr\"odinger equation in terms of the classical Hamiltonian map
does not depend so strongly on the initial conditions. In fact,
these determine only the quantity defined by eq.(\ref{Uo}), whereas
existence of localized/delocalized states is defined entirely by the
localization operator $H(z)$, eq.(\ref{H}), which does not depend at
all on the initial conditions
\cite{Kuzovkov02,Kuzovkov04,Kuzovkov06}. The physical reason for
this is obvious. The ensemble of different trajectories in the AL
can be created already for fixed initial conditions by means of the
ensemble of random potential realizations. For these realizations,
even qualitatively different trajectories in the classical phase
space occur. There is no reason to believe that the disorder
parameter $\sigma$ defines uniquely the trajectory types. It can be
assumed that some random potentials (called \textit{coherent
realization}) correspond to the solutions $\psi_n(\mathbf{k})$ close
to average values $\langle \psi_n(\mathbf{k}) \rangle$ (the
\textit{delocalization} regime). For these realizations excitation
of neighbor modes also occurs, however, at the next discrete-time
steps $n$ these modes return to an equilibrium position,
$\psi_n(\mathbf{k})=0$ (the recurrent behavior). In other words,
only their \textit{virtual} dynamics around equilibrium with limited
amplitude takes place. For other potential realizations (the
\textit{localization} regime) thermalization occurs and the
recurrent behavior disappears. The main physical question -- whether
these two regimes have comparable statistical contributions -- needs
drawing of the phase diagram
\cite{Kuzovkov02,Kuzovkov04,Kuzovkov06}.

Since the detailed AL study in the thermodynamic limit for an
arbitrary space dimension D has already been published by us
\cite{Kuzovkov04}, we restrict ourselves here to the interpretation
of the result from the point of view of the disorder threshold. As
was shown \cite{Kuzovkov04}, for $D \geq D_0=4$  the spectrum of
wave function fluctuations changes, as well as the convergence of
the integral in eq.(\ref{H}). The spatial dimension $D_0=4$ was also
discussed in Ref.\cite{Lee85,Kunz83}, however, the conclusion was
drawn therein that this is an upper critical dimension for
localization, so no phase transitions are expected for $D
> 4$. In other words, no localized states can exist here. Since
appearance of the localized states in the dynamical version of the
Schr\"odinger equation (\ref{Sch}) means \textit{thermalization}
(i.e. energy transfer from initially excited modes to all other
modes), the analog of the above-mentioned result \cite{Kunz83} in
the FPU is a very strong statement on the existence of entirely
recurrent states and no convergence to the thermal equilibrium in
dynamical systems for $D>4$. That is, the Gibbs statistics would not
be applicable for high-dimensional systems. However, this statement
is not based on an exact solutions and looks very suspicious to us.
Thus, let us consider the alternative idea.

According to our Ref.\cite{Kuzovkov04}, for $D \geq D_0=4$  the
problem is fundamentally simplified: for all energies inside the old
band $|E| \leq 2D$ (where only the delocalized states exist) there
is a disorder threshold $\sigma_0(E)$. There is no thermalization
for $\sigma <\sigma_0(E)$, but only delocalized states,
characterized by the (formal) Lyapunov exponent $\gamma\equiv 0$.
For $\sigma \geq \sigma_0(E)$ thermalization occurs, all states are
exponentially localized with a certain $\gamma(\sigma,E)$. As
noticed \cite{Kuzovkov02,Kuzovkov04,Kuzovkov06}, the Lyapunov
exponent $\gamma$ can be treated in the AL problem as the
\textit{long-range order} parameter: the two different phases reveal
different $\gamma$ values, $\gamma \equiv 0$ (delocalized states)
and $\gamma \neq 0$ (localized states, respectively). A sharp
separation in the phase diagram of the localized and delocalized
states recalls the second-order phase transition, where phases
cannot coexist. However, the phase transition here is not of the
second-order, since $\gamma$ reveals step-like changes at the
disorder threshold. As the space dimension D increases, the system
stability increases with respect to disorder, but thermalization in
principle \textit{cannot} completely disappears.

For the case $2 \leq D \leq 3$ (low-dimensionial case
\cite{Kuzovkov04}) the problem becomes more complicated; the
\textit{energy threshold} $E_0=E_0(D)$  also arises. The delocalized
states disappear even under infinitesimal disorder on the boundaries
of the old band, $E_0 \leq |E| \leq 2D$. In other words, only
thermalization is possible in this energy range. There is no analog
of the FPU recurrence here. In the region $|E|\leq E_0$ the disorder
threshold $\sigma_0(E)$ again appears; e.g. for $D=2$ $E_0=2$ and
$\sigma_0(E)=2 (1-\sigma^2/E_0^2)^{1/4}$ \cite{Kuzovkov02}), but its
sense changes. If disorder $\sigma$ exceeds the threshold, only the
localized states exist (full thermalization). As $\sigma <
\sigma_0(E)$,  the disorder parameter $\sigma$ no longer separates
uniquelly the localized and delocalized states, since \textit{phase
co-existence} takes place \cite{Kuzovkov02,Kuzovkov04,Kuzovkov06}.
In other words, in the ensemble of different disorder potential
realizations in this region \textit{both} localized and delocalized
solution can arise with comparable probabilities. As a result of
phase co-existence, phase transition in low-dimensional systems
should be considered as a first-order transition. As was said above,
this non-trivial result has a direct analogy in the statistical
version of the FPU problem.

In the 1D case the analogy between the AL and FPU problems
disappears; the Schr\"odinger equation corresponds here to a
\textit{single} oscillator only and thermalization has no longer a
meaning. Notice, however, that historically the FPU problem arose
during the numerical simulations of a one-dimensional chain of
oscillators \cite{Fermi55,Berman05}. The localization in 1D means
diffusion excitation in one degree of freedom with specific disorder
effects. This is not surprising: when presenting the Schr\"odinger
equation in the form of  eq.(\ref{Sch}), one of the spatial
coordinates in the D-dimensional system was interpreted as a
discrete time variable, respectively, the Hamiltonian dynamics is
treated for oscillator systems in a $p=D-1$ space dimension. That
is, the AL problem in a space dimension D should be compared with
the FPU with a lower space dimension, $p=D-1$.

\subsection{FPU recurrence vs delocalization}

For the sake of illustration, we restrict ourself to the case of the
band centre $E=0$ for 2D system. For this particular case for the
delocalized states ($H(z)=H_{-}(z)$) we calculated analytically
\cite{Kuzovkov02} the inverse Z-transform, eq.(\ref{inverse}), and
found the coefficients  $h_n$ in eq.(\ref{conv}). Eq.(\ref{conv})
can be transformed, respectively:
\begin{eqnarray}\label{conv1}
U_n-U^{(0)}_n = 2\tan (\phi)\sum_{l=1}^{n-1}U^{(0)}_{l} \sin(2 \phi
(n-l)) .
\end{eqnarray}
Eq.(\ref{conv1}) is valid only for $\sigma \leq 2$, provided $2
\sin(\phi)=\sigma$  \cite{Kuzovkov02}.

It is easy to notice that the solution for a disordered system
derives from that for the ordered system with the help of the
sinusoidal modulation. Taking into account that the full squared
coordinate $U^{(0)}_n$  is a sum of oscillating quantities (normal
modes), solution for the perturbed problem $U_n$ is a quasi-periodic
motion. That is, one observes a direct analog of the recurrent
behaviour in the FPU systems.

As was mentioned above, the 2D case for the AL corresponds to the 1D
dynamics. On the other hand, the FPU paradox was established for the
first time namely for the one-dimensional system
\cite{Fermi55,Focus05}, and the result was considered as
counter-intuitive. This is why our exact analytical study of the AL
in the 2D case and the statement on the existence of the delocalized
states \cite{Kuzovkov02} proves once more a close similarity between
the AL and FPU. Indeed, if the recurrent behavior is possible for
the 1D FPU and this result is generally accepted as reliable, there
are no strong grounds to reject our conclusion on the delocalized
states in the 2D AL, since these states are analogous to the 1D
quasi-periodic motion in FPU.

\section{Conclusion}

The Anderson localization and Fermi-Pasta-Ulam problems are very
complementary; each non-trivial and, as a rule, counter-intuitive
result in one problem has its analog in another problem. In
particular, a stable quasi-periodic motion (the recurrent behavior)
in the FPU corresponds to the existence of the delocalized states in
the AL problem. In contrast, the thermalization effect in FPU has an
analog in the localized states in the AL. In general, we have shown
here that a deep analogy between these two problems is possible,
when we treat the quantum tight-binding model in terms of a
classical Hamiltonian map.

This work was partially supported by the Latvian Council of Science.
Author is greatly indebted to  W. von Niessen for detailed
discussions of the paper.


\end{document}